\documentclass[12pt]{article}
\usepackage{epsf}
\begin{document}
\begin{titlepage}

\large
\centerline{\bf WORKSHOP SUMMARY$^{*}$}
\normalsize

\vskip 2.0cm
\centerline{Frederick J. Gilman$^{\dag }$}
\centerline{\it Department of Physics} 
\centerline{\it Carnegie Mellon University} 
\centerline{\it Pittsburgh, Pennsylvania 15213}
\vskip 4.0cm

\centerline{\bf Abstract}
\vskip 1.0cm
Recent progress in understanding the physics of B mesons and 
of CP violation, as presented to this Workshop, is put in historical 
perspective and summarized. 
\vfill

\footnoterule
\noindent $^*$\footnotesize{{\,}Invited talk presented at BCP4, 
Ise-Shima, Japan, February 19-23, 2001; }\\ 
\noindent $^{~}$\footnotesize{{\,}Dedicated to the memory of Abraham Pais, 1919-2000.} \\
\noindent $^\dag$\footnotesize{{\,}Electronic address: gilman@cmuhep2.phys.cmu.edu} 

\end{titlepage}

\newpage

\section{Introduction}

In the middle of the last century it was realized 
that neutral mesons should exist that are not their own antiparticles.  
Such mesons, possessing what we would now call a net flavor quantum 
number, could still `mix' through the weak interactions with their 
respective antiparticles. Therefore, the eigenstates of the resulting 
particle-antiparticle system would be superpositions of particle and 
antiparticle{\,}\cite{GellMannPais}.

The first example of this phenomena was observed in the 1950s for the 
neutral K meson, $K^o$, and its antiparticle, the $\bar{K^o}$.  
Figure 1, from an elegant experiment{\,}\cite{SteinbergerKe3} done in 
the 1970s, shows the result of looking at the proportion of $K^o$'s 
{\em versus} that of $\bar{K^o}$'s as a function of time of flight from 
a production target (where mostly $K^o$'s are produced initially). This 
is measured by looking  at semileptonic decays, for $K^o$ decays result 
in positive leptons and $\bar{K^o}$ decays in negative leptons (here, 
positrons, and electrons, respectively).
\begin{figure}[h]
\vspace{.1in}
\begin{center}
\leavevmode
\epsfxsize=4.5in\epsfbox{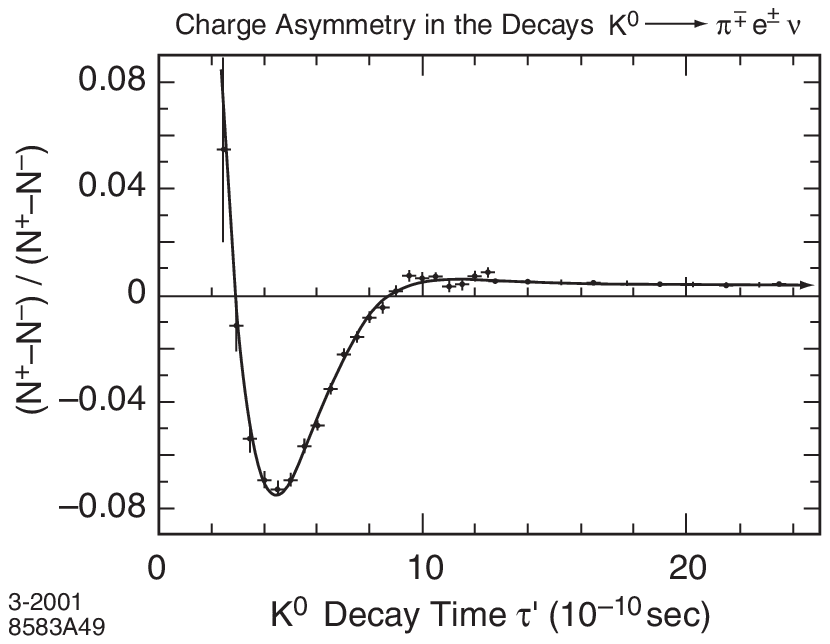}
\end{center}
\caption{The asymmetry in the number of $K^o$ mesons compared to 
the number of $\bar{K^o}$ mesons in a beam as a function of 
time{\,}\cite{SteinbergerKe3}. } 
\label{fig:Ke3asymmetry}
\end{figure}
 
At small times, we can see the preponderance of $K^o$'s.  
The eigenstates, which are mixtures of particle and antiparticle, have 
different masses.  They have widths, and inversely lifetimes, that 
differ by a factor of more than 500, so that we have a  dramatic 
interference pattern as the short-lived ($\tau_S = 0.89 
\times 10^{-10}$~s) $K_S$ dies away, leaving the `long-lived' 
($\tau \approx 5.2 \times 10^{-8}$~s) $K_L$.  
Defining $\Delta M_K = M_{K_L} - M_{K_S}$ and similarly for 
$\Delta \Gamma_K$, it is found that $\Delta M_K = 5.3~{\rm ns}^{-1} 
\approx - \Delta \Gamma_K /2$
from these and other data{\,}\cite{RPP2000}. 

If one looks at times much longer than the $K_S$ lifetime in Figure 1, 
it can be seen that the surviving $K_L$ does not have equal particle and 
antiparticle content: it has slightly more $K^o$ than $\bar{K^o}$. This 
is a manifest breaking of CP symmetry in the neutral $K$ system, usually 
summarized in terms of the parameter $\epsilon$.  At this workshop, a new 
measurement of this charge asymmetry based on 300 million semileptonic 
$K_L$ decays was reported by the KTeV 
Collaboration{\,}\cite{NguyenKTeVKe3}: 
\begin{equation}
{N(e^+ ) - N(e^- ) \over N(e^+ ) + N(e^- ) } = 
{2~{\rm Re}{\,}\epsilon \over 1 + |\epsilon |^2 } = 
3.320 \pm 0.074 \times 10^{-3} ~,
\end{equation}
which is consistent with the previous result{\,}\cite{SteinbergerKe3}, 
but more precise and can be used in conjunction with other 
KTeV results to check conservation of CPT.

CP violation in the $K^o$ system was  of course first 
found{\,}\cite{FitchCronin1964} a decade earlier, not by 
measuring directly the $K^o$ {\em versus} $\bar{K^o}$ content of the $K_L$, 
but by looking at its decays into a CP eigenstate, namely $\pi^+ \pi^-$, 
as shown schematically in Figure 2.

\begin{figure} [h]
\vspace{.1in}
\begin{center}
\leavevmode
\epsfxsize=5in\epsfbox{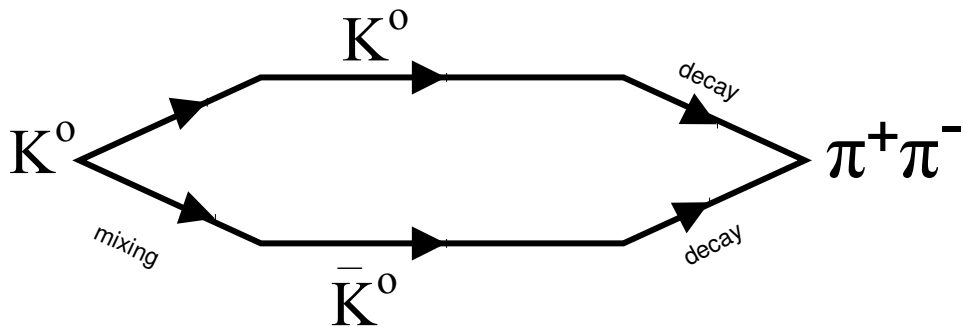}
\end{center}
\caption{The two quantum mechanical paths for an initial $K^o$ 
to decay to a final CP eigenstate consisting of a $\pi^+ \pi^-$.} 
\label{fig:KtoCPeigenstate}
\end{figure}

Starting with an initial $K^o$, we see that because of the possibility 
of mixing there are two quantum mechanical paths to the same final 
state, whose amplitudes must be added.  We take advantage experimentally 
of the large difference of lifetimes in the neutral $K$ system.  
Independent of whether we start with a $K^o$ or $\bar{K^o}$, by waiting 
many $K_S$ lifetimes, nature provides us with the very pure combination 
of $K^o$ and $\bar{K^o}$ that corresponds to a $K_L$ on the right side 
of Figure 2. If CP were conserved, the coherent mixture of $K^o$ and 
$\bar{K^o}$ decay amplitudes would exactly cancel for $K_L \rightarrow 
\pi \pi$.  Instead, the observation that the $K_L$ decays into the 
CP-even eigenstate, $\pi^+ \pi^-$, albeit rarely, shows that CP is 
not conserved and $\epsilon \not= 0$.

\section{${\mbox{\boldmath $B^o$}}$ -- 
${\mbox{\boldmath $\bar{B^o}$}}$ Mixing}

Written in terms of quarks as the fundamental fermionic consituents of 
matter, we recognize the $K^o$ and $\bar{K^o}$ to be the 
$d \bar{s}$ and $s \bar{d}$ combinations of down and strange quarks.  
With the discovery of the charm quark and the bottom quark in the 1970s, 
nature has given us the richness of three additional systems to 
investigate, each with strikingly different properties from the neutral 
K system and from each other.  We will return later to the 
neutral charm mesons, the $D^o = c \bar{u}$ and $\bar{D^o} = u \bar{c}$, 
and focus our attention on the subjects of this Workshop, mesons 
containing a b quark.

The combinations of the down and bottom quarks, $B_d = d \bar{b}$ and 
$\bar{B_d} = b \bar{d}$, form a system where the eigenstates are expected 
and observed to have nearly the same lifetime:  $\Delta \Gamma_d 
<< \Gamma$.  Previous measurements also indicate that 
$\Delta M_d \approx 0.8 \Gamma_d \approx 0.5~{\rm ps}^{-1}$, very close 
to 100 times larger than $\Delta M$ for the neutral K system. Precise 
new measurements of $\Delta M_d$ were presented to this Workshop from 
$B_d - \bar{B_d}$ oscillations of 
\begin{equation}
\Delta M_d = 0.519 \pm 0.020 \pm 0.016~{\rm ps}^{-1} 
\end{equation}
using hadronic decays in BaBAR{\,}\cite{BozziBaBar}, and
\begin{eqnarray}
\Delta M_d &=& 0.463 \pm 0.008~({\rm stat.}) \pm 0.016~({\rm syst.})
~{\rm ps}^{-1} ~, \\ \nonumber
\Delta M_d &=& 0.527 \pm 0.032~({\rm stat.})~{\rm ps}^{-1}~,  
~{\rm and} \\ \nonumber
\Delta M_d &=& 0.522 \pm 0.026~({\rm stat.})~{\rm ps}^{-1} 
\end{eqnarray}
from dileptons, hadronic decays, and semileptonic decays, respectively, 
in Belle{\,}\cite{HastingsBelle}. Some of these are somewhat above the 
previous world average value{\,}\cite{RPP2000} of $0.472 
\pm 0.017~{\rm ps}^{-1}$, and it will be interesting to see how the 
situation develops with further measurements.

\begin{figure} [h]
\vspace{.1in}
\begin{center}
\leavevmode
\epsfxsize=4.5in\epsfbox{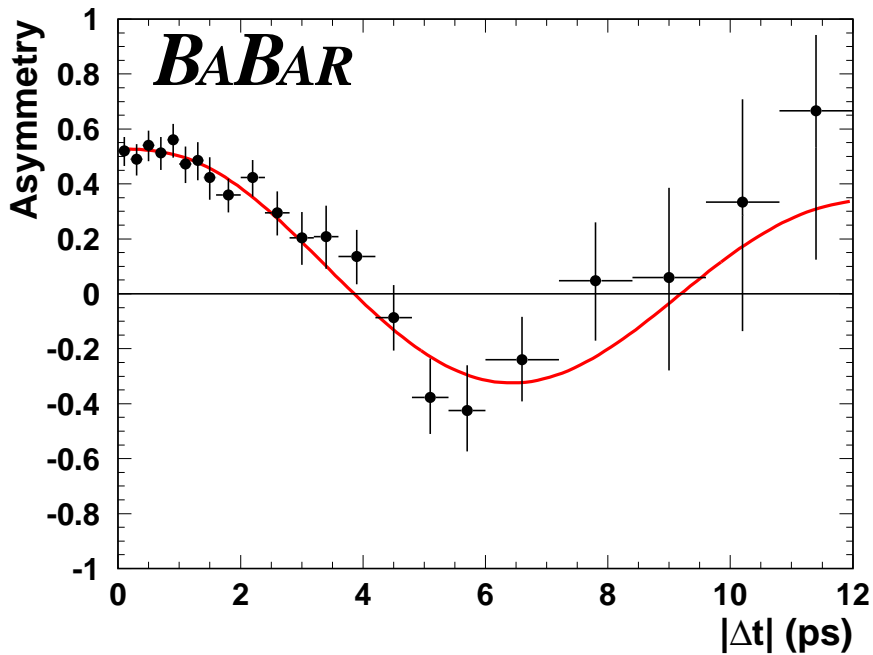}
\end{center}
\caption{The asymmetry in the number of $B_d$ mesons compared to 
the number of $\bar{B_d}$ mesons as a function of time{\,}\cite{BozziBaBar}.} 
\label{fig:BdMixing}
\end{figure}

The full oscillation of an initial $B_d$ to a $\bar{B_d}$ and then back 
to a $B_d$, as presented to this Workshop{\,}\cite{BozziBaBar}, is 
seen in Figure 3.  It is to be contrasted with the oscillation 
seen in the neutral $K$ system in Figure 1 and that of the system 
consisting of the $B_s = s \bar{b}$ and $\bar{B_s} = b \bar{s}$, which 
exhibits features that are different yet again. So far, for the $B_s$ 
we have only the lower limit{\,}\cite{Boix}: 
$\Delta M_s > 15~{\rm ps}^{-1}\approx25 \Gamma_s$ 
from a combination of the LEP and SLD data. Thus, the $B_s$ oscillates to a 
$\bar{B_s}$ and back again at least four times in an average B lifetime! 
In this case we expect{\,}\cite{Hashimoto} a measurable width difference 
$\Delta \Gamma_s \approx 0.1 \Gamma_s$ and the combined 
measurements from CDF and LEP are consistent with this{\,}\cite{Boix}:
\begin{equation}
\Delta \Gamma_s /\Gamma_s = 0.16 {}_{-0.09}^{+0.08} ~.
\end{equation}

While we don't have the large lifetime difference of the neutral K 
system that allowed us to study the particle {\em versus} antiparticle 
content of the $K_L$, we can gain information on the same quantity 
in the neutral B system by measuring the asymmetry between the rates of 
production of two positive leptons compared to two negative leptons 
when a $B^o \bar{B^o}$ pair is produced in electron-positron annihilation 
and both decay semileptonically (after either the $B$ or $\bar{B^o}$ 
mixes).  As in the $K^o$ system, this CP-violating asymmetry is expected 
to be small.  Indeed, at this workshop a new value with considerably 
increased precision, but still consistent with zero, was reported for 
the $B_d$ system by the CLEO Collaboration{\,}\cite{EcklundCLEO}.  When 
combined with a previous analysis that uses both hadronic and semileptonic 
decays, they find that for $\epsilon_B$ defined analogously to 
$\epsilon$ in the neutral K system,  
\begin{equation}
{{\rm Re}{\,}\epsilon_B \over 1 + |\epsilon_B |^2 } = 
0.0035 \pm 0.0103 \pm 0.0015 ~.
\end{equation}

\section{Neutral Meson Mixing }

    At this point, we take a theoretical interlude from the 
procession of beautiful experimental results to recap the 
formalism of mixing in particle-antiparticle systems. 
If we label the meson 
by the index 1 and the anti-meson by the index 2, then the 
time dependence of this two state system is determined by a 
$2 \times 2$ matrix 
\begin{equation}
\left(\matrix{ M_{11} - {i \over 2}\Gamma_{11} & 
               M_{12} - {i \over 2}\Gamma_{12} \cr
		   M_{21} - {i \over 2}\Gamma_{21} &
		   M_{22} - {i \over 2}\Gamma_{22} \cr
		   }\right) ~.
\label{eq:MassMatrix}
\end{equation} 
We have split this $2 \times 2$ matrix into hermitian, $M$, 
and antihermitian, $i\Gamma$, parts, so that:
\begin{eqnarray}
	M_{12} & = & {M_{21}}^{\!*}  \\  \nonumber
	\Gamma_{12} & = & {\Gamma_{21}}^{\!*}  ~,
\label{eq:HermitianMatrix}
\end{eqnarray}
and $M_{11}$, $M_{22}$, $\Gamma_{11}$, and $\Gamma_{22}$
are real.  If we further assume that CPT invariance
holds, then
\begin{eqnarray}
M_{11} & = & M_{22}  = M       \\    \nonumber
\Gamma_{11} & = & \Gamma_{22}  = ~~\Gamma  ~,
\label{eq:CPT}
\end{eqnarray}
and the matrix now reads
\begin{equation}
\left(\matrix{
	M - {i \over 2}\Gamma &
	M_{12} - {i \over 2}\Gamma_{12} \cr
	{M_{12}}^{\!*} - {i \over 2}{\Gamma_{12}}^{\!*} &
	M - {i \over 2}\Gamma \cr
	}\right)     ~.
\label{eq:CPTMassMatrix}
\end{equation}

     The diagonalized matrix gives the mass and width of 
the eigenstates. Furthermore, CP is conserved if and only if 
Im$(M_{12} {\Gamma_{12}}^* ) = 0$. This could happen if $M_{12}$ 
and $\Gamma_{12}$ have the same phase (nearly true for the 
neutral K system), or if one of them vanishes (nearly true 
for $\Gamma_{12}$ in the neutral B systems).

     The large diagonal elements of the matrix get dominant 
contributions from the quark masses and the strong interactions.  
The flavor-changing off-diagonal elements get contributions 
only from the weak interactions. The absorptive off-diagonal 
element, $\Gamma_{12}$, can be related to a sum 
over on-shell intermediate states that couple to both the 
meson and the anti-meson. For the dispersive $M_{12}$, an 
infinite sum of virtual intermediate states of arbitrarily 
increasing mass are relevant.  $M_{12}$ can be computed 
in the Standard Model from weak interaction box diagrams 
involving W's and quarks.  QCD corrections to these weak 
interaction processes are important, but are under good control, 
have been done in leading order and next-to-leading order.  
At the vertices of these box diagrams are the elements of 
the Cabibbo-Kobayashi-Maskawa matrix, to which we now turn 
our attention.

\section{The Cabibbo-Kobayashi-Maskawa Matrix}

In the Standard Model with $SU(2) \times U(1)$ as the gauge 
group of electroweak interactions, both the quarks and leptons are 
assigned to be left-handed doublets and right-handed singlets. 
The quark mass eigenstates differ from the weak eigenstates, and
the matrix relating these bases was defined for six quarks and 
given an explicit parametrization by 
Kobayashi and Maskawa{\,}\cite{KobayashiMaskawa73} in 1973.  
It generalizes the four-quark case, where the matrix is parametrized 
by a single angle, the Cabibbo angle{\,}\cite{Cabibbo63}.

By convention, the mixing is often expressed in terms of a 
$3\times 3$ CKM matrix $V$ operating on the charge $-e/3$ 
quark mass eigenstates ($d$, $s$, and $b$):
\begin{equation}
\left(\matrix{d ^{\,\prime}   \cr
                s ^{\,\prime} \cr
                b ^{\,\prime} \cr
       		}\right)
=
	\left(\matrix{
		V_{ud}&     V_{us}&     V_{ub}\cr
		V_{cd}&     V_{cs}&     V_{cb}\cr
		V_{td}&     V_{ts}&     V_{tb}\cr
       		}\right)
	\left(\matrix{
		d \cr
                s \cr
                b \cr
       		}\right) ~.
\end{equation}

As this matrix V is unitary, it can physically be fully specified by 
four real parameters.  These can be taken to be three ``rotation'' 
angles and one phase.  CP violation has a natural place and occurs if 
the phase is not $0^o$ or $180^o$ and the other angles are not $0^o$ 
or $90^o$, {\em i.e.}, if there is non-trivial mixing between each 
pair of generations of quarks and there is a non-trivial phase.  

     Decades of experimental effort have produced a great deal of 
information about the magnitude of the CKM matrix elements from 
semileptonic decays, neutrino reactions, and most recently, 
hadronic $W$ decays{\,}\cite{GKRreview2000}. We know experimentally 
that all three angles that characterize the CKM matrix are small 
but non-zero.  In the absence of a well-motivated argument to 
the contrary, there is an expectation that the single non-trivial 
phase should be non-zero as well.  If CP violation does arise from 
the CKM matrix, it gives rise to both a natural scale and a 
special pattern for CP-violating effects (and 
for flavor-changing-neutral-current effects generally).

\section{The Unitarity Triangle}

Direct and indirect information on the less precisely known 
elements of the CKM matrix is neatly summarized in terms of the 
``unitarity triangle,'' one of six such triangles that correspond 
to the unitarity condition applied to two different rows or 
columns of the CKM matrix. Unitarity applied to the first and 
third columns yields
\begin{equation}
V_{ud} ~{V_{ub}}^{\!*} + V_{cd} ~{V_{cb}}^{\!*} 
+ V_{td} ~{V_{tb}}^{\!*} = 0 ~.
\label{eq:FullUnitarityTriangle}
\end{equation}

The unitarity triangle is just a geometrical presentation of this 
equation in the complex plane. We can 
always choose to orient the triangle so that $V_{cd} ~{V_{cb}}^*$~
lies along the horizontal or real axis, as it is in a number of 
standard parametrizations.  Setting diagonal elements of the 
CKM matrix to unity (which is good to a few percent or better) 
and recognizing that $V_{cd} = s_{12}$, the sine of the Cabibbo 
angle, Eq.~(\ref{eq:FullUnitarityTriangle}) becomes
\begin{equation}
{V_{ub}}^{\!*} + V_{td} = s_{12} ~{V_{cb}}^{\!*} ~.
\label{eq:UnitarityTriangle}
\end{equation}
Rescaling the triangle by a factor $[1/|s_{12} ~V_{cb}|]$
so that the base is of unit length, 
the coordinates of the vertices become 
\begin{equation}
    A\bigl( \hbox{Re}(V_{ub})/|s_{12} ~V_{cb}|,
-\hbox{Im}(V_{ub})/|s_{12}\,V_{cb}| \bigr) ~, B(1,0) ~, C(0,0) ~.
\label{eq:ScaledUnitarityTriangle}
\end{equation}
In the Wolfenstein parametrization\cite{Wolfenstein83},
the coordinates of the vertex $A$ of the rescaled unitarity 
triangle are simply $(\rho, \eta)$.

A non-trivial unitarity triangle, given the other information 
that we have on the magnitudes of the CKM matrix elements, 
implies that there is CP violation in the Standard Model and 
{\em vice versa}.  Since we know the length of the base, 
$|V_{cb} s_{12} |$, quite well and that of another side,  
$|V_{ub} |$, moderately well, the complete triangle could be 
determined by either a measurement of the length of the 
third side $|V_{td} |$, one of the angles, or by some condition 
that involves a combination of these.  

Ultimately we want to obtain multiple determinations of the 
triangle to test the Standard Model as a description of weak 
interactions and of CP violation.  A failure of the multiple 
determinations to agree would lead us to our even more 
important goal of getting clues to the physics that lies 
beyond the Standard Model.  

The traditional inputs to determination of the unitarity triangle 
are $|V_{cb} |$ and $|V_{ub} |$ from semileptonic B 
decays, plus, under the assumption that they originate in the 
Standard Model, $\epsilon$ from the neutral K system, $\Delta M_d$ 
from $B_d$ mixing, and (limits on) $\Delta M_s$ from $B_s$ mixing.  
Reviews of earlier measurements{\,}\cite{FortyLEP} of $|V_{cb} |$ 
and $|V_{ub} |$, as well as new 
measurements{\,}\cite{EcklundCLEO},{\,}\cite{JangBelle},
{\,}\cite{ThorndikeCLEO} were presented to this Workshop.  
In particular, a different type of 
analysis{\,}\cite{ThorndikeCLEO} from CLEO that uses moments in 
inclusive B semileptonic and $b \rightarrow s \gamma$ decays to 
fix the heavy-quark-effective-theory parameters gives another 
way of determining $|V_{cb} |$ with error bars that are at 
least comparable to other methods and should lead to a more 
precise value of $|V_{ub} |$ in the near future. 

The enterprise of trying to pin down the unitarity triangle 
has been underway for more than a decade, and great progress 
has been made.  It is instructive to go back a dozen years 
to when the B-factories were just being put forward to measure the 
angles of the unitarity triangle. Not only were the uncertainties 
at that time considerably larger on the magnitudes of the CKM 
relevant matrix elements, but because the top mass was not 
known, the constraints from $\epsilon$ and B-mixing had to be 
applied with the added uncertainty of a big potential range for $m_t$.  
This is illustrated{\,}\cite{DDGN1989} in Figure 4, 
where the case of $m_t = 160$ GeV has been picked out from many 
other figures in 1989 that covered a then-perceived range of 
80 to 200 GeV for $m_t$.  
\pagebreak
\begin{figure} [h]
\vspace{.1in}
\begin{center}
\leavevmode
\epsfysize=5in\epsfbox{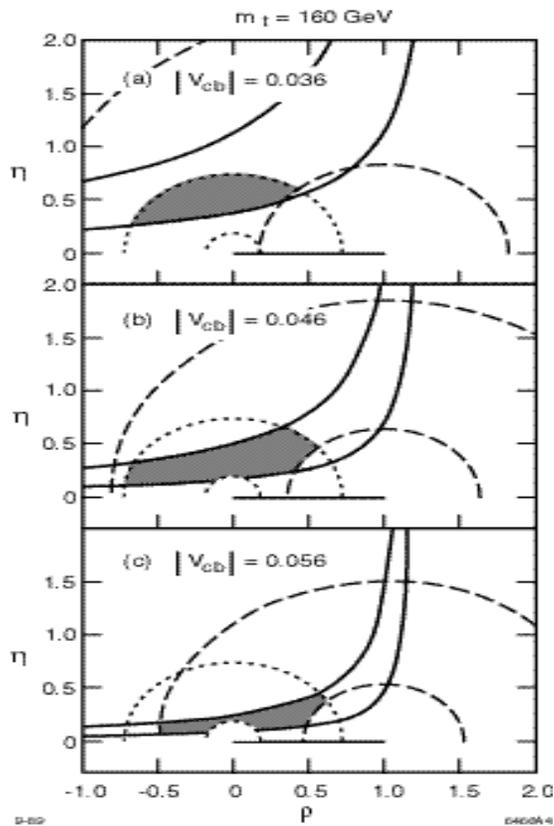}
\end{center}
\caption{The restrictions{\,}\cite{DDGN1989} on the position of the 
vertex A of the unitarity triangle in the $\rho$ - $\eta$  
plane in 1989 for a top quark mass, $m_t = 160$~GeV.} 
\label{fig:UnitarityTriangle1989}
\end{figure}

Compare this with the more recent status{\,}\cite{GKRreview2000} 
of the unitarity triangle shown in Figure 5.  By not dividing 
by the length of the base, we emphasize the accuracy ($\sim 10^{-3}$ 
of the magnitude of the diagonal elements of the matrix) that is 
now needed to make further progress.  We are indeed entering an 
era of precision CKM measurements.  

\begin{figure}[ht]
\vspace{.1in}
\begin{center}
\leavevmode
\epsfxsize=5in\epsfbox{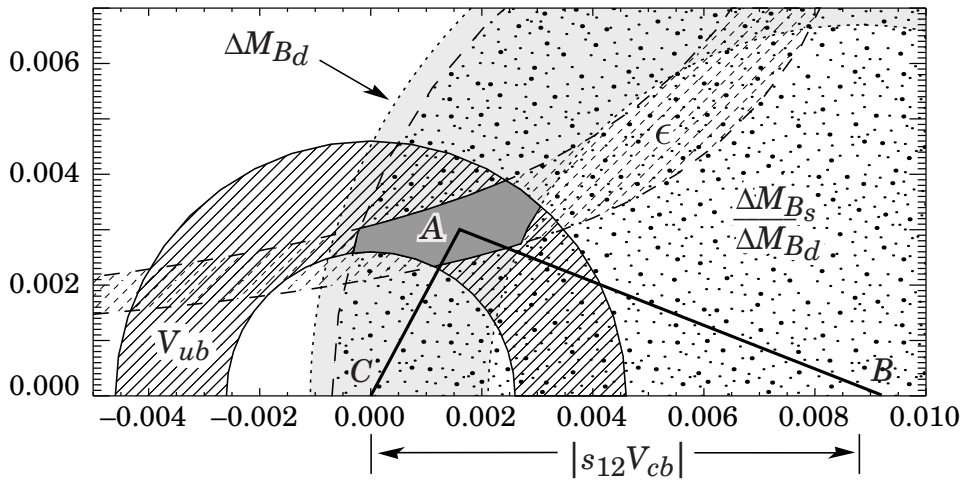}
\end{center}
\caption{Constraints on the position of the vertex, 
A, of the unitarity triangle following from $|V_{ub} |$,
$B$ mixing, and $\epsilon$ are indicated. A possible 
unitarity triangle is shown with A in the preferred 
region{\,}\cite{GKRreview2000}. }
\label{fig:UnitarityTriangleConstraints}
\end{figure}

The biggest uncertainties remaining are theoretical in character. 
They lie particularly in the hadronic matrix elements of operators 
containing quark fields.  There are intensive discussions underway, 
both of the size of these theoretical errors and of how to treat them. 
This includes a number of recent papers and contributions to this 
Workshop{\,}\cite{Lacker},{\,}\cite{FalkCKM},{\,}\cite{HockerCKM},{\,}
\cite{Soni}. It was the subject of a panel at the B2000 meeting as 
well{\,}\cite{B2000panel}.  While this may appear to be a controversy 
merely evoking strongly stated positions among a small segment of 
the community, it has important physics consequences{\,}\cite{B2000panel}. 
Taking smaller uncertainties in $|V_{ub}|$ and the hadronic parameters 
involved in $B_d$ mixing (to extract $|V_{td} |$), leads some to 
claim{\,}\cite{B2000panel},{\,}\cite{Ciuchini2000} that there is a 
strong likelihood that we must have a non-trivial triangle, {\em i.e.}, 
that there is CP violation arising from the CKM matrix, without 
imposing any constraint from $\epsilon$ in the neutral K system 
(whose use would already be assuming that CP violation originates 
in the CKM matrix).  Others{\,}\cite{HockerCKM},{\,}\cite{B2000panel}, 
including me, are not ready to draw such a strong conclusion yet.  
Continued progress is being 
made{\,}\cite{Hashimoto},{\,}\cite{Aoki},{\,}\cite{Kronfeld} in 
lattice QCD calculations to determine the hadronic parameters.  
Ultimately, they should produce results with an 
accuracy that is comparable to those from the experimental 
measurements themselves. 

Where will the next major advance come in this area?  I believe that 
it will come with the measurement of $B_s$ mixing at the Tevatron 
collider.  The prospects{\,}\cite{Taylor} for such a measurement 
were described at this Workshop.  It appears that if the mixing  
originates through Standard Model physics, $B_s$ mixing will be 
measured rather accurately, and that will allow the comparison with 
$B_d$ mixing to give $|V_{ts} /V_{td} |$ with small theoretical 
uncertainties. Since $|V_{ts} |$ must be very close to $|V_{cb} |$ 
with three generations, this in turn will fix $|V_{td} |$ in the 
Standard Model with small errors and determine the unitarity triangle 
from measurements of the three sides.  Then we will see if this is 
consistent with the other information we will have on the angles 
or the area of the triangle.

\section{CP Violation Involving ${\mbox{\boldmath $B^o$}}$ Mixing}

The decay of an initial $B^o$ to a final state, $f$, can occur 
though an amplitude corresponding to $B^o \rightarrow f$, or 
by the $B^o$ mixing to a $\bar{B}^o$, followed by the $\bar{B}^o$ 
decaying to the same final state through an amplitude corresponding 
to $\bar{B}^o \rightarrow f$.  This gives rise to the 
situation{\,}\cite{BigiSanda} shown in Figure 6.

\begin{figure}[h]
\vspace{.1in}
\begin{center}
\leavevmode
\epsfxsize=5in\epsfbox{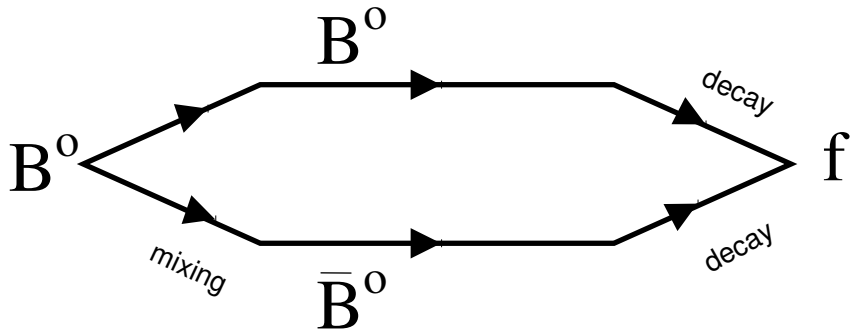}
\end{center}
\caption{Two quantum mechanical paths for $B^o$ decay through 
mixing and decay to a final state $f$. } 
\label{fig:BMixingDecay}
\end{figure}

On the one hand, note the close theoretical similarity to the 
situation involving mixing and decay in the neutral $K$ system that we 
discussed earlier.  On the other hand, because of the lifetime of the 
$B$, instead of the hundreds of meters from production to decay that 
contemporary $K_L$ decay experiments involve, at an $e^+ e^-$ 
B-factory experiments must be able to measure the distance from 
the production of a $B$ to its decay of hundreds of microns -- 
roughly a million times smaller! 

If $f$ is a CP eigenstate, $f_{\rm CP}$, and we take the neutral 
$B$ eigenstates to have the same lifetime, then the time-dependent 
rate for an initial $B^o$/$\bar{B}^o$ to decay into $f_{\rm CP}$ is 
\begin{equation}
{d\Gamma [B^o (t)/\bar{B}^o (t) \rightarrow f_{\rm CP} ]\over dt} = 
e^{-\Gamma t} ~[1 \pm \eta_{\rm CP} \sin 2\phi 
\sin (\Delta M t) ]~, 
\end{equation}
where $\eta_{\rm CP}$ is the CP-eigenvalue of the final state and 
the phase $\phi$ arises from CKM matrix elements relevant to the 
particular decay.  In simple cases where one weak amplitude 
dominates, $\phi$ turns out to be just an angle of the unitarity 
triangle. These angles are taken to be $\alpha = \phi_2$, 
$\beta = \phi_1$, and $\gamma = \phi_3$ at the vertices A, B, and 
C, respectively, of the unitarity triangle.

\section{CP Violation in B Decay Amplitudes}

One can also have CP-violating observables that do not involve 
mixing.  They occur through the interference of two decay 
amplitudes that contribute to a given process and its CP-conjugate 
process.  Under charge conjugtion, the weak phases change 
sign, while the strong phases do not, as both C and P separately 
are conserved by the stong interactions.  To get a non-zero rate 
difference, one must have at least two amplitudes with different 
weak and strong phases. In the $B_d$ and $B_s$ systems, there are 
simple relations between CP-violating rate differences for processes 
obtained by interchanging d and s quarks{\,}\cite{GronauReview}.

An example is provided by the neutral $K$ system, where 
interference of tree and penguin ampitudes, which have different 
weak phases, for the decay $K \rightarrow \pi \pi$ give a non-zero 
rate difference characterized by the parameter $\epsilon^\prime$, 
with{\,}\cite{RPP2000} 
$\epsilon^\prime / \epsilon \sim 2 \times 10^{-3}$. 
Many possibilities for such CP-violating asymmetries can 
be found in $B$ decays, but their observation remains for the 
future.

\section{The B-factories} 

Before looking at the new experimental results on CP violation in the 
neutral $B$ system, one must salute the members of the accelerator 
physics community that designed and built the asymmetric electron-positron 
colliding beam machines, the B-factories, that allow these experiments 
to be done at all.  The performance of PEPII at SLAC{\,}\cite{SeemanPEPII} 
and of KEKB at KEK{\,}\cite{KurokawaKEKB} have been absolutely remarkable.  
The peak luminosity of PEPII has exceeded the design goal of $3 \times 
10^{33}$~${\rm cm}^{-2}$${\rm s}^{-1}$ and KEKB was operating above 
$2.4 \times 10^{33}$~${\rm cm}^{-2}$${\rm s}^{-1}$ during the Workshop.  
A dozen years ago, there were plenty of skeptics that one could have 
amperes of positrons and electrons of different energies colliding 
to produce physics results.  But today we sit solidly in the range of 
integrated luminosities that were foreseen{\,}\cite{DDGN1989} at 
that time as needed to produce a statistically significant measurement 
of a CP-violating asymmetry.

This success has given us planning toward realizing what 
would once have seemed truly amazing possibilities.  PEPII is 
on an upgrade path{\,}\cite{SeemanPEPII} that gets to luminosities 
of $5 \times 10^{33}$ this year and to $10^{34}$ in 2003.  
Feasibility studies are being conducted of several avenues to 
$3 \times 10^{34}$~${\rm cm}^{-2}$${\rm s}^{-1}$.  
KEKB{\,}\cite{KurokawaKEKB} plans 
call for reaching $7 \times 10^{33}$~${\rm cm}^{-2}$${\rm s}^{-1}$ 
in 2002-2003 and more than $10^{34}$ in 2005-2006.  These 
luminosities inspired an exciting session{\,}\cite{Alexander10**34} 
of this Workshop, ``$10^{34}$ and Beyond,'' to examine the 
experimental possibilities that are opened up by such luminosities.

\section{CP Violation in the B System: \\
Measurement of 
${\mbox{\boldmath $\sin 2\beta = \sin 2\phi_1$}}$  }

The BaBar and Belle Collaborations have both presented new data 
to this Workshop on tagged ${B_d}^o$ and $\bar{B_d}^o$ decays 
to final CP eigenstates, $f_{\rm CP}$, to observe a time-dependent 
CP asymmetry{\,}\cite{Smith2beta},{\,}\cite{Hazumi2beta}. 
Both employ the ``golden'' final state $J/\psi K_S$, plus 
$\psi(2S) K_S$ and $J/\psi K_L$.  Belle uses a few other modes 
as well.  Through the time-dependent asymmetry, these decays all 
provide a measure of $\sin 2\beta = \sin 2\phi_1$ 
from the unitarity triangle in a theoretically clean way.  
Furthermore, the high-performance subsystems of both detectors 
all work so as to make this prime measurement for which 
they were designed. 

The results are 
\begin{eqnarray}
\sin 2\beta   &=& 0.34 \pm 0.20 \pm 0.05 
~({\rm BaBar}{\,}\cite{Smith2beta}) \\
\sin 2\phi_1 &=& 0.58 ~~{}_{-~0.34}^{+~0.32} 
~~{}_{-~0.10}^{+~0.09} ~~({\rm Belle}{\,}\cite{Hazumi2beta})
\end{eqnarray}
These results are consistent with each other and with previous 
measurements.  The present world average is 
\begin{equation}
\sin 2\beta = 0.48 \pm 0.16
\end{equation}
\pagebreak

\begin{figure}[h]
\vspace{.1in}
\begin{center}
\leavevmode
\epsfysize=4in\epsfbox{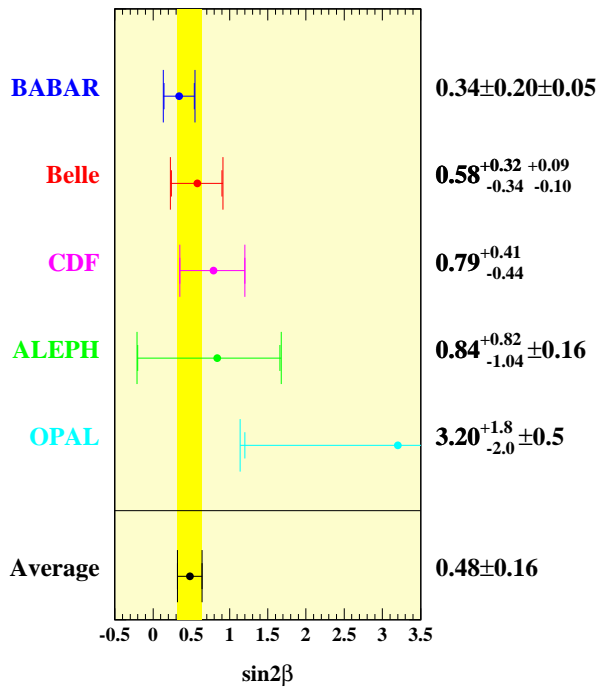}
\end{center}
\caption{The world's measurements of $\sin 2\beta$ from  
CP-violating asymmetries in neutral $B$ mixing and 
decay{\,}\cite{Smith2beta}. } 
\label{fig:WorldSine2beta}
\end{figure}

Thus, combining all the data, it is already unlikely that CP 
is conserved in the combination of neutral $B$ mixing and decay. 

These results are also consistent with the Standard Model and a 
single phase from the CKM matrix as the origin of CP violation.  
For example, from Figure 5, $\sin 2\beta$ should lie in the 
range from 0.5 to 0.9.   

We can now see getting to the goal of errors of 0.1 or less for 
the ``golden''mode, $J/\psi K_S$, and having confirmatory measurements 
in several other modes.  However, at the present, the asymmetry 
data for just the final state $J/\psi K_S$ with the 
$K_S$ decaying to $\pi^+ \pi^-$ are (with statistical errors only) 
$0.25 \pm 0.26$ and $1.21 {}_{-0.47}^{+0.40}$ from BaBar and Belle, 
respectively.  We need a little more patience to get to the decisive 
test for which we have been waiting many years.

\section{Measurement of the Other Angles of the Unitarity Triangle}

Early on, it seemed that $\sin 2\beta$ would be the 
first of the CP-violating asymmetries to be measured.  This has 
indeed turned out to be true, as we have just seen.  The other 
angles of the unitarity triangle have not only proven to be 
harder than expected to get at experimentally, but they have 
also turned out to be theoretically more complicated as well.

For $\sin 2\alpha = \sin 2\phi_2$, the mode that attention was 
originally focused on was $B^o /\bar{B}^o \rightarrow \pi^+ \pi^-$.  
Not only has the branching ratio for this mode turned out to be 
considerably smaller than had been hoped (see the next section), 
but it was that in addition to a tree amplitude, penguin 
diagrams with different weak phases likely enter in a significant 
way.  Much theoretical work that depends on measurements of other 
two body modes plus $SU(2)$ and/or $SU(3)$ symmetry has gone into 
finding ways to salvage the situation.  The measurement of other 
decay modes such as $\sin 2\alpha$ could prove important as well. 
Removing this ``penguin polution'' to re-establish a clean 
measurement of $\sin 2\alpha$ has been reviewed both at this 
Workshop{\,}\cite{GronauReview} and elsewhere{\,}\cite{BABARbook}.  

For $\sin 2\gamma = \sin 2\phi_3$, many methods have been 
proposed{\,}\cite{GronauReview},{\,}\cite{BABARbook}.  
Two recent ideas requiring very high luminosities were the subject 
of talks{\,}\cite{Falk2gamma},{\,}\cite{Atwood2gamma} at this 
Workshop.  During the course of the Workshop I inquired of a 
number of my theoretical colleagues as to where the first rough 
measurement of $\gamma$ will come from.  The consensus was that 
most likely a comparison{\,}\cite{GronauRosnerLondon} 
of accurately measured branching ratios 
in $B \rightarrow K \pi$ decays (where there is an interference 
of tree and penguin amplitudes that involves the weak phase 
$\gamma$) would yield the first information directly restricting 
the range of gamma.  Eventually, a precise value was seen to come 
from $B_u \rightarrow D^o K$ rates at a high luminosity 
electron-positron collider{\,}\cite{Alexander10**34}, {\,}\cite{BABARbook} 
or from time-dependent $B_s (t) \rightarrow D_s K$ measurements at 
hadron colliders.  While the latter measurements might begin during 
Run II of the Tevatron{\,}\cite{Taylor}, they are primarily the domain 
of the physics program of the next 
generation of hadron collider experiments, BTEV{\,}\cite{JohnsBTEV} 
and LHCb{\,}\cite{NakadaLHCb}.

\section{Rare B Decays}

The measurement of many branching ratios for B decays at the 
$10^{-5}$ level and below is proceeding apace at BaBar, Belle, 
and CLEO.  This Workshop had more than half a dozen talks on 
measurements of rare B decays{\,}\cite{KawasakiBelle},{\,}
\cite{HockerBaBar},{\,}\cite{IijimaBelle},{\,}\cite{LyonCLEO},
{\,}\cite{UshirodaBelle},{\,}\cite{BozekBelle},{\,}
\cite{GarmashBelle} plus papers in the poster sessions,  
and more than that number of theoretical talks examining how 
to understand the mechanism  by which such decays 
proceed{\,}\cite{GronauReview},{\,}\cite{Mannel},{\,}\cite{Li},
{\,}\cite{Keum},{\,}\cite{Silvestrini},{\,}\cite{London},{\,}
\cite{Lipkin},{\,}\cite{Brodsky} and how new physics might show 
up{\,}\cite{Masiero},{\,}\cite{Hou},{\,}\cite{Berger}, 
plus more papers in the poster sessions.  I only have space 
to call attention to a few of the many interesting developments 
reported here. 

\begin{itemize}

\item{The measurements of the branching ratios for $B$ decays 
to two pseudoscalar mesons have settled down and now we have 
substantial agreement between BaBar{\,}\cite{HockerBaBar}, 
Belle{\,}\cite{IijimaBelle}, and CLEO{\,}\cite{LyonCLEO}.  
In particular, for the branching ratio for 
$B^o \rightarrow \pi^+ \pi^-$ they  report 
$4.1 \pm 1.0 \pm 0.7 \times 10^{-6}$,  
$5.9 {}_{-0.21}^{+0.24} \pm 0.5 \times 10^{-6}$, and  
$4.3 {}_{-1.4}^{+1.6} \pm 0.5 \times 10^{-6}$, respectively. 
As noted above, even aside from theoretical difficulties, 
this low branching ratio makes deriving information on 
$\sin 2\alpha$ that much more difficult.  }

\item{Systematic studies of all the ``Cabibbo-suppressed''
$B \rightarrow D K^-$ and $B \rightarrow D^* K^-$ decays shows 
that their branching ratios are at the expected level of 
$\sim 0.07$ times their pionic counterparts{\,}\cite{IijimaBelle}. }

\item{Inclusive and exclusive measurements of B decays that 
correspond to $b \rightarrow s \gamma$ are becoming standard 
measurements{\,}\cite{ThorndikeCLEO},{\,}\cite{UshirodaBelle}, 
and we can look forward soon to measurements of 
$b \rightarrow s \mu^+ \mu^-$. The corresponding theoretical 
situation is in rather good shape{\,}\cite{Mannel}. }

\item{Important theoretical progress is being made as well, 
centering on understanding QCD as it applies to B non-leptonic 
decays{\,}\cite{Li},{\,}\cite{Keum},{\,}\cite{Silvestrini},
{\,}\cite{london},{\,}\cite{Lipkin},{\,}\cite{Brodsky}.  Specific 
questions are where and how factorization applies and what is the 
importance of penguin diagrams in the overall picture.  These issues 
were the focus of the presentation{\,}\cite{Silvestrini} on factorization 
and penguin amplitudes in $B \rightarrow K \pi$ decays, where 
precise data should be forthcoming that will clarify whether we 
now have a good theoretical understanding of these and similar decays.}

\end{itemize}

\section{D, K, and Lepton Physics}

     Our review of what is happening and will happen in B physics 
is not complete without taking into account the complementary 
experimental and theoretical work for other quark and lepton flavors 
which relate to the same theoretical parameters and questions.  As 
pointed out in an excellent review at this Workshop{\,}\cite{BigiCharm}, 
charm physics, aside from its own merits, acts as a staging area both 
experimentally and theoretically for the assault on B physics.  
In addition, it may give us some surprises of its own regarding new 
physics. The particular 
places to watch are the meaurements of $x= \Delta M/\Gamma$ and 
$y = \Delta \Gamma/\Gamma$ in the neutral D system and on CP violating 
asymmetries, as we push to the few percent level described at this 
Workshop{\,}\cite{SmithCLEO},{\,}\cite{TanakaBelle} and smaller in 
the years to come. 

    Experiments involving K mesons continue to provide alternate 
measurements of rare and CP-violating processes that allow 
theoretically clean determinations the CKM parameters.  The future 
programs of KTEV{\,}\cite{ArentonKTEV} and NA48{\,}\cite{KochNA48} 
were described at this Workshop, and are part of a bigger worldwide 
effort that aims at measuring branching ratios, such as that for 
the ``golden mode, '' $K_L \rightarrow \pi^o \nu \bar{\nu}$, 
down to the $10^{-11}$ level{\,}\cite{KAON99}.

   Finally, there are the measurements of the magnetic dipole moment 
of the muon and the search for (T-violating) electric dipole moments. 
We heard about an improved limit on the electric dipole moment of 
the tau{\,}\cite{CoanCLEO} at this Workshop and through another 
talk{\,}\cite{Duong:g-2} shared in the excitement generated by the 
recent measurement of the magnetic moment of the muon, which is 
in tantalizing disagreement with Standard Model predictions.

\section{Conclusion}

We have reached the time where the colliders, detectors, and experimental 
collaborations are in place to carry out the long-planned exploration of 
CP violation in the B system.  Theory has made considerable progress as 
well in understanding how to relate the measurements that will be made 
in the next few years to the fundamental parameters of the theory, 
multiple routes of varying experimental difficulty and theoretical 
cleanliness to those parameters, and in understanding the effects 
that various types of physics beyond the Standard Model could have on 
CP-violating effects in the B system. 

At this Workshop, exciting new results from BaBar and Belle on the 
CP-violating asymmetry that corresponds to $\sin 2\beta = \sin 2\phi_1$ 
have been presented.  We can now see that rather precise measurements of 
the $\beta = \phi_1$ will be made in the next couple of years.  Those, 
together with rough information on the other angles of the unitarity 
triangle and precise magnitudes for the CKM matrix elements that will 
come in the same time frame will make for a decisive test of whether the 
Standard Model picture for the weak interactions between quarks 
and CP violation is correct.

The B-Factory and hadron collider experiments are also exploring a host 
of related issues, from precise measurements of mixing in the neutral 
B systems to rare, flavor-changing-neutral-current decays and CP 
violation without mixing in B decays.  So prepare for enormous 
amounts of data and great physics!

\section*{Acknowledgements} 

I thank Tony Sanda and all the local organizers of BCP4 for their 
excellent arrangements and enhancing the interplay of theory and 
experiment at an unusually opportune time. This research work is 
supported in part by the U.S. Department of Energy under Grant 
No. DE-FG02-91ER40682.


\begin{thebibliography} {99}
\bibitem{Alexander10**34}J. Alexander, summary talk 
on the  ``$10^{34}$ and Beyond'' session, 
and contributions to that session, these Proceedings.
\end{thebibliography}
\end{document}